\documentstyle[aps,epsf,preprint]{revtex}
\begin{document}
\draft
\title{Radiative decays of \boldmath{$\phi$}-meson and
nature of light scalar resonances}
\author{N.N. Achasov}
\address{Laboratory of Theoretical Physics,
 Sobolev Institute for Mathematics,
  Novosibirsk, 630090, Russia}
  \date{\today}
 \maketitle
\begin{abstract}
Based on gauge invariance, we show that the new threshold
phenomenon is discovered in the $\phi$ radiative decays
$\phi\to\gamma a_0\to\gamma\pi^0\eta$ and
 $\phi\to\gamma f_0\to\gamma\pi^0\pi^0$. This enables to conclude
 that production of the lightest scalar mesons $a_0(980)$ and
 $f_0(980)$ in these decays
  is caused by the four-quark transitions, resulting in strong restrictions on the
 large $N_C$ expansions of the decay amplitudes. The analysis
 shows that these constraints give new evidences in favor
 of the four-quark nature of $a_0(980)$ and $f_0(980)$ mesons.
\end{abstract}
  The discovered more than thirty years ago the lightest scalar mesons
$a_0(980)$ and $f_0(980)$ became the hard problem for the naive
quark-antiquark ($q\bar q$) model from the outset. Really, on the
one hand the almost exact degeneration of the masses of the
isovector $a_0(980)$ and isoscalar $f_0(980)$ states revealed
seemingly the structure similar to the structure of the vector
$\rho$ and $\omega$ mesons , and on the other hand the strong
coupling of $f_0(980)$ with the $K\bar K$ channel pointed
unambiguously to a considerable part of the  strange quark pair
$s\bar s$ in the wave function of $f_0(980)$.

In 1977 R.L. Jaffe  noted that in the MIT bag model, which
incorporates confinement phenomenologically, there are light
four-quark scalar states \cite{jaffe}. He suggested that
$a_0(980)$  and $f_0(980)$   might be these states with symbolic
structures $a^0_0(980)=(us\bar u\bar s - ds\bar d\bar s)/\sqrt{2}$
and $f_0(980)=(us\bar u\bar s + ds\bar d\bar s)/\sqrt{2}$. From
that time $a_0(980)$ and $f_0(980)$ resonances came into beloved
children of the light quark spectroscopy, see, for example,
reviews \cite{montanet,achasov-84}.

Ten years later we showed \cite{achasov-89} that the study of the
radiative decays $\phi\to\gamma a_0\to\gamma\pi\eta$ and
$\phi\to\gamma f_0\to \gamma\pi\pi$ can shed light on the problem
of  $a_0(980)$ and $f_0(980)$ mesons. Over the next ten years the
question was considered from different points of view
\cite{bramon,achasov-97,achasov-97a,achasov-97b}.

Now these decays have been studied not only theoretically but also
experimentally, so that the time is ripe to sum up.

Present time data have already been obtained from Novosibirsk with
the detectors SND
 \cite{snd-fit,snd-ivan} and CMD-2 \cite{cmd},
 which give the following
branching ratios :
$BR(\phi\to\gamma\pi^0\eta)=(0.88\pm0.14\pm0.09)\cdot10^{-4}$
\cite{snd-fit} (2000), $BR(\phi\to\gamma\pi^0\pi^0)=
(1.221\pm0.098\pm0.061)\cdot10^{-4}$ \cite{snd-ivan} (2000) and
$BR(\phi\to\gamma\pi^0\eta)=(0.9\pm0.24\pm0.1)\cdot10^{-4}$,
$BR(\phi\to\gamma\pi^0\pi^0)=(0.92\pm0.08\pm0.06)\cdot10^{-4}$
\cite{cmd}. DA$\Phi$NE also confirms the Novosibirsk results
\cite{barbara}.

 These data give evidence in favor of the four-quark $(q^2\bar
q^2)$ \cite{jaffe,achasov-84,black,josef} nature of  $a_0(980)$
and $f_0(980)$ mesons \cite{achasov-89,achasov-97,achasov-01}.
Note that the isovector $a_0(980)$ meson is produced in the
radiative $\phi$ meson decay
 as intensively as the well-studied $\eta '(958)$ meson containing $\approx 66\%$ of $s\bar s$,
  responsible for the decay .
 It is a clear qualitative argument
for the presence of the $s\bar s$ pair in the isovector $a_0(980)$
state, i.e., for its four-quark nature.

Since the one-loop model  $\phi\to K^+K^-\to\gamma a_0$ and
$\phi\to K^+K^-\to\gamma f_0$, suggested at basing the
experimental investigations \cite{achasov-89}, was used in the
data treatment from the outset, the question on the mechanism of
the scalar meson production in the $\phi$ radiative decays was put
into the shade. We show below that the present data give the
conclusive arguments in favor of the $K^+K^-$ loop mechanism of
$a_0(980)$ and $f_0(980)$ mesons production in the $\phi$
radiative decays. The knowledge of this mechanism allows to
conclude that the production of $a_0(980)$ and $f_0(980)$ in the
$\phi$ radiative decays is caused by the four-quark transitions.
This constrains the large
 $N_C$ expansions of the decay amplitudes and gives new impressive evidences in favor
 of the four-quark nature of $a_0(980)$ and $f_0(980)$.

\begin{figure}
\centerline{\epsfxsize=14cm \epsfysize=9cm \epsfbox{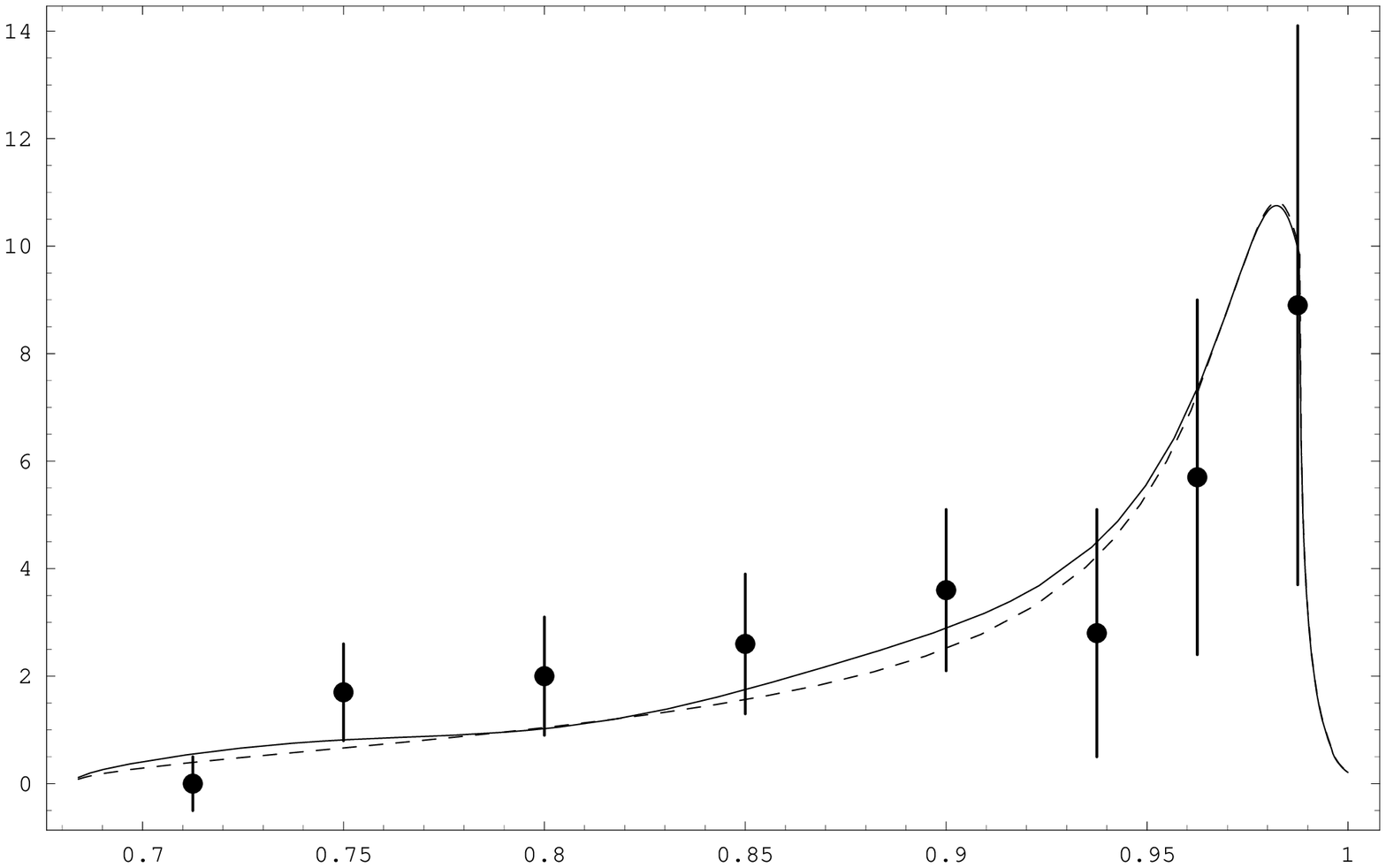}}
 \caption{ Fitting of $dBR(\phi\to\gamma\pi^0\eta\,,\,m)/dm\times 10^4\mbox{GeV}^{-1}$  is shown with the solid line,
 the signal contribution $\phi\to\gamma a_0\to\gamma\pi^0\eta$ is shown with the dashed line.}  \label{figa0}
\end{figure}

\begin{figure}
\centerline{\epsfxsize=14cm \epsfysize=9cm
\epsfbox{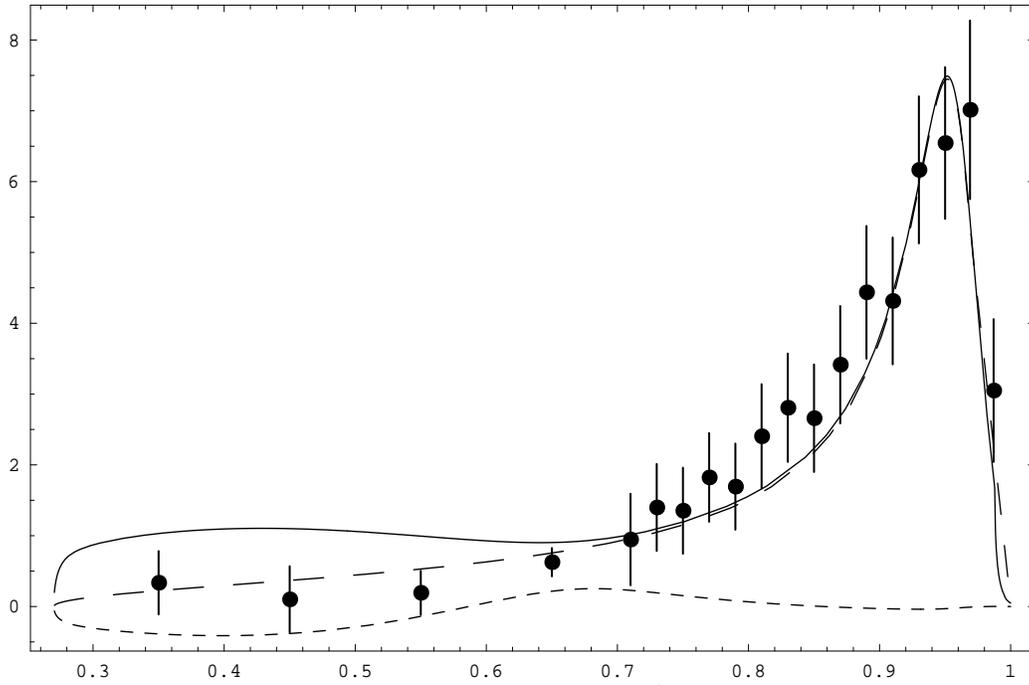}}
 \caption{
Fitting of $dBR(\phi\to\gamma\pi^0\pi^0\,,\,m)/dm\times
10^{4}\mbox{GeV}^{-1}$ is shown with the solid line, the signal
contribution $\phi\to\gamma f_0\to\gamma\pi^0\pi^0$ is shown with
the dashed line. The dots show the interference.} \label{figf0}
\end{figure}

In Figs. \ref{figa0} and \ref{figf0} are shown the SND data on
$\phi\to\gamma\pi^0\eta$ \cite{snd-fit} (2000) and
$\phi\to\gamma\pi^0\pi^0$ \cite{snd-ivan} (2000). The data are
described in the following model $\phi\to(\gamma
a_0+\pi^0\rho)\to\gamma\pi^0\eta$ and $\phi\to(\gamma
f_0+\pi^0\rho)\to\gamma\pi^0\pi^0$, see details in Ref.
\cite{achasov-01}. It is important for us now nothing but the fact
that the $\phi\to\gamma a_0\to\gamma\pi^0\eta$ process dominates
everywhere over the region of the $\pi^0\eta$ invariant mass
$m_{\pi^0\eta}= m$ and the $\phi\to\gamma f_0\to\gamma\pi^0\pi^0$
process dominates in the resonance region of the $\pi^0\pi^0$
system, the $\pi^0\pi^0$ invariant mass $m_{\pi^0\pi^0}=m > 670$
MeV \cite{cut}.

The resonance mass spectrum is of the form \cite{mixing}

\begin{eqnarray}
&&S_R(m)=d\Gamma(\phi\to\gamma R\to\gamma ab\,,\, m)/dm\nonumber\\
&&=\frac{2}{\pi}\frac{m^2\Gamma(\phi\to\gamma R\,,\,m)\Gamma(R\to
ab\,,\,m)}{|D_R(m)|^2} = \frac{4|g_R(m)|^2\omega (m)
p_{ab}(m)}{3(4\pi)^3m_{\phi}^2}\left |\frac{g_{Rab}}{D_R(m)}\right
|^2,
 \label{spectrumR}
\end{eqnarray}
where $R = a_0\ \mbox{or}\ f_0$ and $ab=\pi^0\eta\ \mbox{or}\
\pi^0\pi^0$ respectively, $\omega (m)=(m_{\phi}^2-m^2)/2m_{\phi}$
is the photon energy in the $\phi$ meson rest frame, $p_{ab}(m)$
is the modulus of the $a$ or $b$ particle momentum in the $a$ and
$b$ mass center frame, $g_{Rab}$ is the coupling constant,
$g_{f_0\pi^0\pi^0}=g_{f_0\pi^+\pi^-}/\sqrt{2}$, $D_R(m)$ is the
$R$ resonance propagator the form of which everywhere over the $m$
region can be find in \cite{achasov-89,achasov-80,achasov-01a},
$g_R(m)$ is the invariant amplitude that describes the vertex of
the $\phi (p)\to\gamma (k) R(q)$ transition with $q^2=m^2$. This
is precisely the function which is the subject of our
investigation.

By  gauge invariance, the transition amplitude is proportional to
the electromagnetic field strength tensor $F_{\mu\nu}$ (in our
case to the electric field in the $\phi$ meson rest frame):
\begin{eqnarray}
&& A\left [\phi(p)\to\gamma (k) R(q)\right ]= G_R(m)\left [p_\mu
e_\nu(\phi) - p_\nu e_\mu(\phi)\right]\left [k_\mu e_\nu(\gamma) -
k_\nu e_\mu(\gamma)\right],
 \label{gauge}
\end{eqnarray}
 where  $e(\phi)$ and
$e(\gamma)$ are the $\phi$ meson and $\gamma$ quantum polarization
four-vectors, $G_R(m)$ is the invariant amplitude free from
kinematical singularities. Since there are no charge particles or
particles with magnetic moments in the process, there is no pole
in $G_R(m)$. Consequently, the function
\begin{eqnarray}
g_R(m)= - 2(pk)G_R(m) = - 2\omega (m) m_\phi G_R(m) \label{gRm}
\end{eqnarray}
is proportional to $\omega (m)$ (at least!) in the soft photon
region.

To describe the experimental spectra in Figs. \ref{figa0} and
\ref{figf0}, the function $|g_R(m)|^2$ should be smooth (almost
constant) in the range $m\leq 0.99$ GeV, see Eq.
(\ref{spectrumR}). Stopping the function $(\omega (m))^2$ at
$\omega (990\,\mbox{MeV})=29$ MeV with the help of the form-factor
$1/\left [1+(R\omega (m))^2\right ]$ requires $R\approx 100$
GeV$^{-1}$. It seems to be incredible to explain the formation of
such a huge radius in hadron physics. Based on the large, by
hadron physics standard, $R\approx10$ GeV$^{-1}$, one can obtain
an effective maximum of the mass spectrum under discussion only
near 900 MeV. To exemplify this trouble let us consider the
contribution of the isolated R resonance: $g_R(m)=-2\omega (m)
m_\phi G_R\left (m_R\right )$. Let also the  mass and the width of
the R resonance equal 980 MeV and 60 MeV, then
$S_R(920\,\mbox{MeV}):S_R(950\,\mbox{MeV}):S_R(970\,\mbox{MeV}):S_R(980\,\mbox{MeV})
=3:2.7:1.8:1$.

So stopping the $g_R(m)$ function is the crucial point in
understanding  the mechanism  of the production of  $a_0(980)$ and
$f_0(980)$ resonances in the $\phi$ radiative decays.

\begin{figure}
\centerline{\epsfxsize=14cm \epsfysize=8.5cm \epsfbox{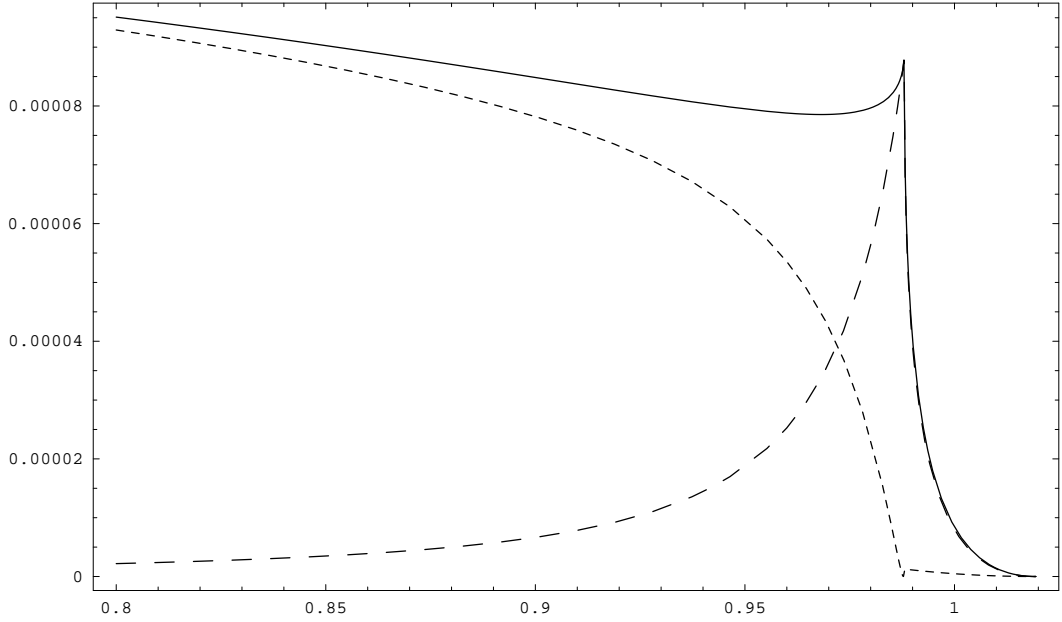}}
 \caption{ The function $|g(m)|^2$ is drawn with the solid line. The contribution of the
  imaginary part is drawn with the dashed line. The contribution of the real part
   is drawn with the dotted line.} \label{g}
\end{figure}

The $K^+K^-$ loop model $\phi\to K^+K^-\to\gamma R$
\cite{achasov-89} solves this problem in the elegant way: the fine
threshold phenomenon is discovered, see Fig. \ref{g}, where the
universal in $K^+K^-$ loop model function $|g(m)|^2=\left
|g_R(m)/g_{Rab}\right |^2$ is shown \cite{gm}.

\begin{figure}
\centerline{\epsfxsize=14cm \epsfysize=8.5cm \epsfbox{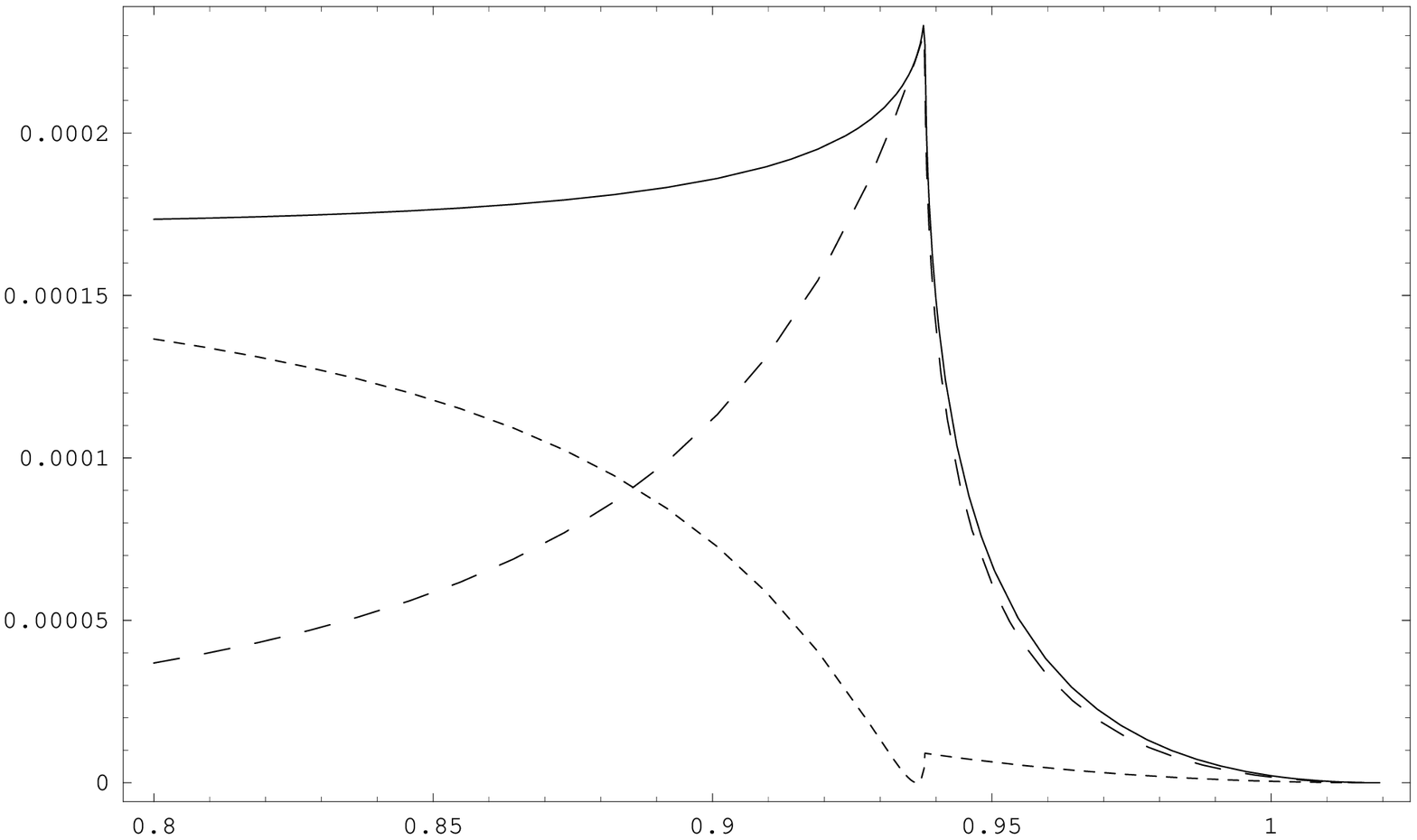}}
 \caption{ The function $|g(m)|^2$ for  $m_{K^+}=469$ MeV is drawn
 with the solid line. The contributions of the imaginary and real
 parts are drawn with the dashed and dotted lines respectively.
 }
  \label{gg}
\end{figure}

\begin{figure}
\centerline{ \epsfxsize=14cm \epsfysize=8.5cm \epsfbox{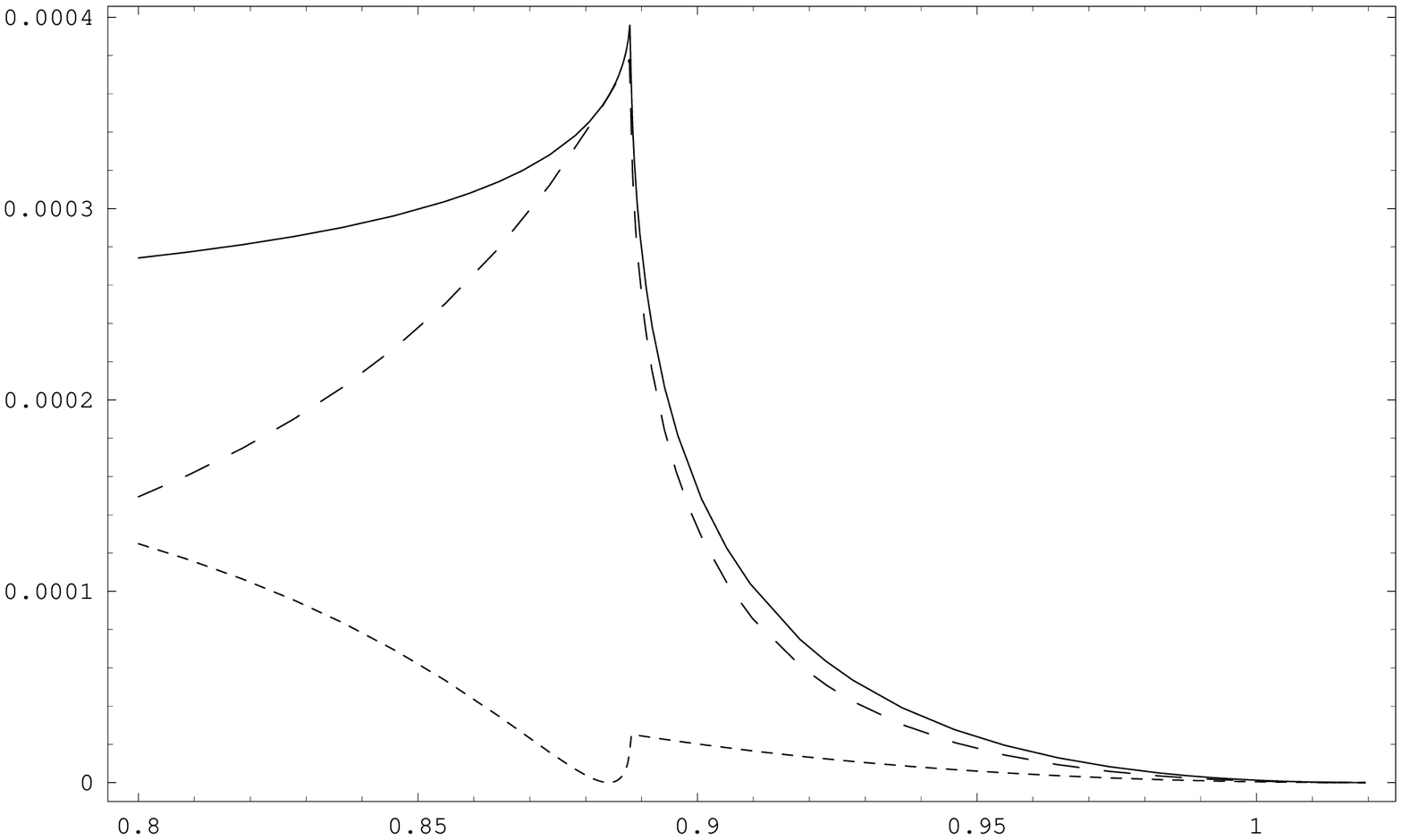}}
 \caption{ The function $|g(m)|^2$ for  $m_{K^+}=444$ MeV is drawn
  with the solid line. The contributions of the imaginary and real
   parts are drawn with the dashed and dotted lines respectively.
  } \label{ggg}
\end{figure}

 To demonstrate the threshold character of this effect we present
 Fig. \ref{gg} and Fig. \ref{ggg} in which the function $|g(m)|^2$
is shown in the case of $K^+$ meson mass is 25 MeV and 50 MeV less
than in reality.

One can see from Figs. \ref{gg} and \ref{ggg} that the function
$|g(m)|^2$ is suppressed by the $(\omega (m))^2$ law in the region
950-1020 MeV and 900-1020 Mev respectively \cite{phone}.

In the mass spectrum this suppression is increased by one more
power of $\omega (m)$, see Eq. (\ref{spectrumR}), so that we
cannot see the resonance in the region 980-995 MeV. The maximum in
the spectrum is effectively shifted to the region 935-950 MeV and
880-900 MeV respectively.

In truth this means that $a_0(980)$ and $f_0(980)$ resonances are
seen in the radiative decays of $\phi$ meson owing to the $K^+K^-$
intermediate state, otherwise the maxima in the spectra would be
shifted to 900 MeV.

So the mechanism of production of $a_0(980)$ and $f_0(980)$
mesons in the $\phi$ radiative decays is established.

Both the real part and the imaginary one of the $\phi\to\gamma R$
 amplitude are caused by the $K^+K^-$ intermediate state. The
imaginary part is caused by the real $K^+K^-$ intermediate state
while the real part is caused by the virtual compact $K^+K^-$
intermediate state, i.e., we are dealing here with the four-quark
transition \cite{im} . Needless to say, radiative four-quark
transitions can happen between two $q\bar q$ states as well as
between $q\bar q$ and $q^2\bar q^2$ states but their intensities
depend strongly on a type of the transitions. A radiative
four-quark transition between two $q\bar q$ states requires
creation and annihilation of an additional $q\bar q$ pair, i.e.,
such a transition is forbidden according to the Okuba-Zweig-Izuka
(OZI) rule, while a radiative four-quark transition between $q\bar
q$ and $q^2\bar q^2$ states requires only creation of an
additional $q\bar q$ pair, i.e., such a transition is allowed
according to the OZI rule.

Let us consider this problem from the large $N_C$ expansion
standpoint. Let us begin with the $q\bar q$ model.

In the two-quark model $a^0_0(980)=(u\bar u - d\bar d)/\sqrt{2}$,
the large $N_C$ expansion of the $\phi = s\bar s\to\gamma
a_0(980)$ amplitude starts with the OZI forbidden transition of
the $1/N_C$ order ( $s\bar s$ annihilation and $u\bar u,\, d\bar
d$ creation). But its weight is bound to be small, because this
term does not contain the $K^+K^-$ intermediate state emerging
only in the next to leading term of the $\left (1/N_C\right )^2$
(!) order for creation and annihilation of additional $q\bar q$
pairs. It will be recalled that $\phi\to\gamma\eta ' (958)$
transition does not require creation of an additional $q\bar q$
pair at all (the OZI superallowed transition) and has the $\left
(N_C\right )^0$ order.

In the two-quark model $f_0(980)=(u\bar u + d\bar d)/\sqrt{2}$,
which involves the $a_0$\,-$f_0$ mass degeneration, the large
$N_C$ expansion of the $\phi = s\bar s\to\gamma f_0(980)$
amplitude starts also with the OZI forbidden transition of the
$1/N_C$ order, whose weight also is bound to be small, because
this term does not contain the $K^+K^-$ intermediate state
emerging only in the next to leading term of the $\left
(1/N_C\right )^2$ order.

In the two-quark model $f_0(980)=s\bar s$, which has the serious
trouble with the $a_0$\,-$f_0$ mass degeneration,  the $\left
(N_C\right )^0$ order transition without creation of an additional
$q\bar q$ pair ( similar in this regard to $\phi\to\gamma\eta '
(958)$ ) is bound to have a small weight in the large $N_C$
expansion of the $\phi = s\bar s\to\gamma f_0(980)$ amplitude,
because this term does not contain the $K^+K^-$ intermediate state
emerging only in the next to leading term of the $1/N_C$ order,
i.e., in the OZI forbidden transition. Emphasize that the
mechanism without creation  and annihilation of the additional
$u\bar u$ pair  cannot explain the $S_{f_0}(m)$ spectrum because
it does not contain the $K^+K^-$ intermediate state!

But if $a^0_0(980)$ and $f_0(980)$ mesons are compact $K\bar K$
states $(u\bar ss\bar u - d\bar ss\bar d)/\sqrt{2}$ and $(u\bar
ss\bar u + d\bar ss\bar d)/\sqrt{2}$
 respectively, i.e., four-quark states similar to states of
the MIT-bag model \cite{molecule}, the large $N_C$ expansions of
the $\phi = s\bar s\to\gamma a_0(980)$ and $\phi = s\bar
s\to\gamma f_0(980)$ amplitudes start  with the OZI allowed
transitions of the $\left (1/N_C\right )^{-1/2}$ order, which
require only creation the additional $u\bar u$ pair for the
$K^+K^-$ intermediate state. It will be recalled that a OZI
allowed hadronic decay amplitude, for example, the $\rho\to\pi\pi$
amplitude, has the $\left (1/ N_C\right ) ^{-1/2}$ order.

In summary the fine threshold phenomenon is discovered, which is
to say that the $K^+K^-$ loop mechanism of
 the $a_0(980)$ and $f_0(980)$ scalar meson production in the
$\phi$ radiative decays is established at a physical level of
proof. The case is rarest in hadron physics. This production
mechanism is the four-quark transition what constrains the large
$N_C$ expansion of the $\phi\to\gamma a_0(980)$ and $\phi\to\gamma
f_0(980)$ amplitudes and gives the new strong (if not crucial)
evidences in favor of the four-quark nature of $a_0(980)$ and
$f_0(980)$ mesons.

\begin{references}
\bibitem{cut}
The difference from the experimental data, observed in the region
$m_{\pi\pi}<670$ MeV, is due to the fact that in the experimental
processing of the $e^+e^-\to\gamma\pi^0\pi^0$ events the
background events $e^+e^-\to\omega\pi^0\to\gamma\pi^0\pi^0$ are
excluded with the help of the invariant mass cutting, in so doing
the part of the $e^+e^-\to\phi\to\rho\pi^0\to\gamma\pi^0\pi^0$
events is excluded as well.
\bibitem{mixing}
Notice that in Ref. \cite{achasov-01} we took into account the
mixing of $f_0(980)$ meson with another scalar isoscalar
resonance, see also Ref. \cite{achasov-97}, but such a
complication is not essential for the present investigation.
\bibitem{gm}
The forms of $g_R(m)$ and $g(m)=g_R(m)/g_{Rab}$ everywhere over
the $m$ region are in Refs. \cite{achasov-89} and
\cite{achasov-01a} respectively.
\bibitem{im}
It will be recalled that the imaginary part of every hadronic
amplitude describes a multi-quark transition.
\bibitem{molecule}
In the case of the $K\bar K$ bound states with the binding energy
close to 20 MeV , i.e.,in the extended molecule case, the
contribution of the virtual intermediate $K^+K^-$ states in the
$K^+K^-$ loop is suppressed by the momentum distribution in the
molecule, and the real part of the loop amplitudes are negligible
\cite{achasov-97a}. It leads to the branching ratios much less
than the experimental ones. In addition, the $S_R(m)$ spectra in
the $K\bar K$ molecule case are much narrower than the
experimental ones, see the behavior of the imaginary part
contribution in Fig. \ref{g}.
\end{references}
\end{document}